%% file: article.tex
\documentclass[12pt]{article}
\usepackage{epsfig}

\input econfmacros.tex

\def\CP      {\ensuremath{C\!P}\ }
\def\phm {\ensuremath{\phantom{-}}}

\def\Title#1{\begin{center} {\Large {\bf #1} } \end{center}}

\begin{document}

\Title{Study of direct \CP in charmed $B$ decays and 
measurement of the CKM angle $\gamma$ at Belle}

\bigskip\bigskip

\begin{raggedright}  

{\it Karim Trabelsi\index{Trabelsi, K.}\\
KEK (High Energy Accelerator Research Organization),\\
1-1 Oho, Tsukuba, Ibaraki,\\
305-0801, JAPAN}
\bigskip\bigskip
\end{raggedright}

\begin{abstract}
The Belle experiment, running at the KEKB $e^+ e^-$ asymmetric energy
collider during the first decade of the century, has recorded 
770 fb$^{-1}$ of data at the $\Upsilon(4S)$ resonance. A combination
of recent Belle results obtained with this sample is used to perform 
a measurement of the CKM angle $\gamma$. We use $B^\pm \to DK^\pm$ and 
$B^\pm \to D^*K^\pm$ decays where the $D$ meson ($D^0$ or ${\overline{D}}^0$) 
decays into $K_S^0\pi\pi$, $K\pi$, $KK$, $\pi\pi$, $K_S^0 \pi^0$ and 
$K_S^0 \eta$ final states and $D^*$ decays into $D\pi^0$ and $D\gamma$. 
Belle obtains the most precise $\gamma$ measurement to date, 
$\gamma = (68 {}^{+15}_{-14})^\circ$.
\end{abstract}
\vskip7.0cm
\begin{center}
Proceedings of CKM 2012, \\
the 7th International Workshop on the CKM Unitarity Triangle, \\
University of Cincinnati, USA, 28 September - 2 October 2012 
\end{center}
\newpage

\section{Introduction}
\par
\begin{figure}[hpt]
\begin{center}
\includegraphics[width=0.65\textwidth,clip]{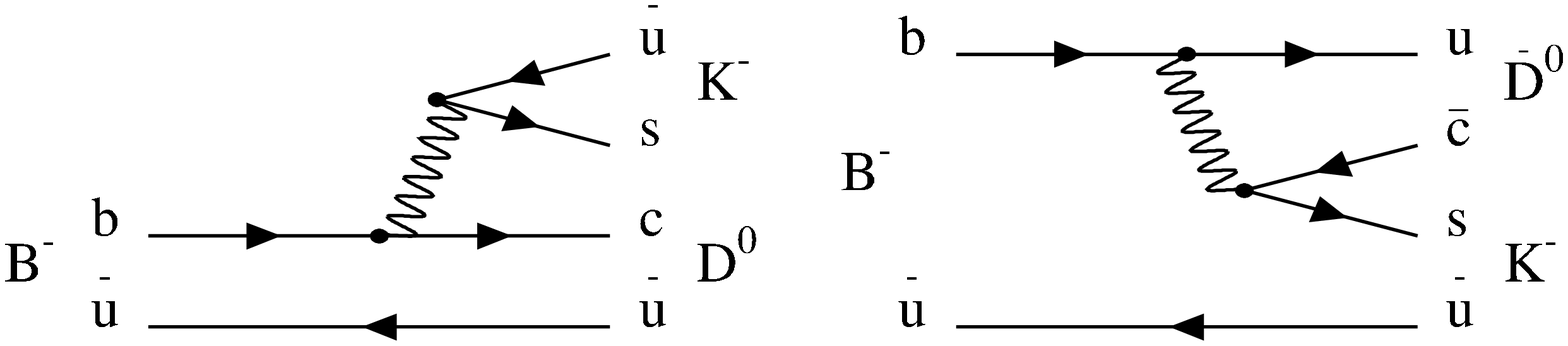}
\end{center}
\caption{Feynman diagrams for $B^- \to D^0 K^-$ and  $B^- \to        
{\overline{D}}^0 K^-$. }
\label{feynman}
\end{figure}

 Two angles of the CKM unitarity triangle, $\beta$ and $\alpha$, have 
now been measured with high precision~\cite{hfag}. 
The determination of the third angle, $\gamma$, using $B^\pm \to DK^\pm$ decays,
will require much more data than for the other angles. 
Its determination 
is however theoretically clean due to the absence of loop contributions; 
$\gamma$ can be determined using tree-level processes only, exploiting the
interference between $b \to c \overline{u} d$ and $b \to u \overline{c} d$
transitions that occurs when a process involves a neutral $D$ meson
reconstructed in a final state accessible to both $D^0$ and ${\overline{D}}^0$
decays (Fig~\ref{feynman}). Therefore, the angle $\gamma$ provides a SM 
benchmark, and its precise measurement is crucial in order to disentangle 
non-SM contributions to other processes, via global CKM fits.
The size of the interference also depends on the ratio ($r_B$) 
of the magnitudes of the two tree diagrams involved
and $\delta_B$, the strong phase difference between them. Those hadronic
parameters will be extracted from data together with the angle $\gamma$.
The value of $r_B$ is the product of the ratio of the CKM matrix elements
$|V_{ub}^* V_{cs}|/|V_{cb}^* V_{us}| \sim 0.38$ and the color suppression
factor, which result in a value of around 0.1, whereas
$\delta_B$ can not be precisely calculated from theory. 
Note that $r_B$ and $\delta_B$ can take different values for different 
$B$ decays: the values of $B^\pm \to DK^\pm$ and $B^\pm \to D^*K^\pm$ are not the same.
Several different $D$ decays have been studied in order to maximize the
sensitivity to $\gamma$. The archetype is the use of $D$ decays to \CP
eigenstates, a method proposed by M.~Gronau, D.~London, and D.~Wyler 
(and called the GLW method)~\cite{GLW}.
%The difficulties in the application of the GLW method arise primarily
%due to the small magnitude of the \CP asymmetry of the $B^\pm \to D_{CP}K^\pm$
%decay, which may lead to significant systematic uncertainties
%in the observation of the \CP violation. 
An alternative approach was
proposed by D.~Atwood, I.~Dunietz, and A.~Soni~\cite{ADS}. 
Instead of using $D^0$ decays to \CP eigenstates, 
the ADS method uses Cabibbo-favored and doubly
Cabibbo-suppressed $D$ decays. In the decays $B^+ \to [K^-\pi^+]_D K^+$ and 
$B^- \to [K^+\pi^-]_D K^-$, the suppressed $B$ decay is followed by a
Cabibbo-allowed $D^0$ decay, and vice versa. Therefore, the interfering
amplitudes are of similar magnitude, and one can expect a large 
\CP asymmetry. The main limitation of the method is that the branching 
fractions of those decays above are small. 
A Dalitz plot analysis of a three-body $D$ meson final state allows
one to obtain all the information required for the determination of $\gamma$
in a single decay mode. Three-body final states such as $K_S^0 \pi^+ \pi^-$
have been suggested as promising modes~\cite{GGSZ} for the extraction of 
$\gamma$.
At present the Dalitz method (or GGSZ method) has the best sensitivity to 
$\gamma$.

Latest Belle results using the full data sample taken at the $\Upsilon(4S)$
(corresponding to $772 \times 10^6$ $B\overline{B}$ pairs) 
are described in these proceedings and the resulting $\gamma$,
combining these different results, is given.

\section{GGSZ results}

As in the GLW and ADS methods, the two amplitudes interfere if the $D^0$ and
$\overline{D}^0$ mesons decay into the same final state $K_S^0 \pi^+ \pi^-$. 
Assuming no \CP asymmetry in neutral
$D$ decays, the amplitude for $B^+ \to D[K_S\pi^+\pi^-]K^+$ decay 
as a function of Dalitz plot variables 
$m_+^2 = m^2_{K_S^0 \pi^+}$ and $m_-^2 = m^2_{K_S^0 \pi^-}$ is 
\begin{equation}
f_{B^+} = f_D (m^2_+, m^2_-) + r_B e^{i\gamma + i\delta_B} f_D (m^2_-, m^2_+)
\end{equation}
where $f_D (m^2_+, m^2_-)$ is the amplitude of the $\overline{D}^0 \to
K_S^0 \pi^+ \pi^-$ decay.
Similarly, the amplitude for $B^- \to D[K_S\pi^+\pi^-]K^-$ decay 
is 
\begin{equation}
f_{B^-} = f_D (m^2_-, m^2_+) + r_B e^{-i\gamma + i\delta_B} f_D (m^2_+, m^2_-).
\end{equation}
The $\overline{D}^0 \to K_S^0 \pi^+ \pi^-$ decay amplitude $f_D$ can be
determined from a large sample of flavor-tagged 
$\overline{D}^0 \to K_S^0 \pi^+ \pi^-$ decays produced 
in the continuum $e^+ e^-$ annihilation. Once $f_D$ is
known, a simultaneous fit to $B^+$ and $B^-$ data allows the contributions
of $r_B$, $\gamma$ and $\delta_B$ to be separated. The method has only 
two-fold ambiguity: $(\gamma, \delta_B)$ and $(\gamma + 180^\circ, \delta_B
+ 180^\circ)$ solutions cannot be distinguished. 
Due to the fact that $r_B$ is bound to be positive, the direct extraction
of $r_B$, $\delta_B$ and $\gamma$ can be biased. To avoid these biases, the
Cartesian coordinates have been introduced, 
$x^\pm = r_B \cos (\delta_B \pm \gamma)$ and
$y^\pm = r_B \sin (\delta_B \pm \gamma)$.
A combined unbinned maximum likelihood fit to the $B^+$ and $B^-$ samples
with free parameters $(x^\pm,\ y^\pm)$ yields the 
values given in Table~\ref{tab_ggsz}.
Combining $B^\pm \to D K^\pm$ and $B^\pm \to D^* K^\pm$, the value 
$\gamma = (78 {}^{+11}_{-12} \pm 4 \pm 9)^\circ$ is obtained~\cite{Belle_GGSZ},
where the quoted uncertainties are respectively statistical, systematic and 
due to an imperfect knowledge of the amplitude model that describes 
$D \to K_S^0 \pi^+ \pi^-$ decays. 
\begin{table}[htb]
\caption{Results of Belle GGSZ analyses.}
\label{tab_ggsz}
\begin{center}
\begin{tabular}{lcc}
\hline \hline
Observables &  $B \to DK$ & $B \to D^\star K$ \\
\hline
$x^+$ & $-0.107 \pm 0.043 \pm 0.011$ & $+0.083 \pm 0.092 \pm 0.011$ \\
$y^+$ & $-0.067 \pm 0.059 \pm 0.018$ & $+0.157 \pm 0.109 \pm 0.018$ \\
$x^-$ & $+0.105 \pm 0.047 \pm 0.011$ & $-0.036 \pm 0.127 \pm 0.011$ \\
$y^-$ & $+0.177 \pm 0.060 \pm 0.018$ & $-0.249 \pm 0.118 \pm 0.018$ \\
%Mode &  $\gamma$ $({}^\circ)$ & $\delta_B$  $({}^\circ)$ &  $r_B$ \\
%\hline
%$B \to DK$ & $81 {}^{+13}_{-15} \pm 5$ & 
%$137 {}^{+13}_{-16} \pm 4$ & $0.16 \pm 0.04 \pm 0.01$ \\
%$B \to D^\star K$ & $74 {}^{+19}_{-20} \pm 4$ & $342 {}^{+19}_{-21} \pm 3$
%& $0.20 \pm 0.07 \pm 0.01$ \\
\hline \hline
\end{tabular}
\end{center}
\end{table}
The last source of uncertainty can be eliminated by binning the Dalitz 
plot (Refs.~\cite{GGSZ, GGSZ_2}),
using information on the average strong phase difference between $D^0$ and
${\overline{D}}^0$ decays in each bin that can be determined using the
quantum-correlated $\psi(3770)$ data. Such results have been published recently
by CLEO-c~\cite{CLEOc}. The measured strong phase difference is used 
to obtain the model-independent result~\cite{Belle_GGSZ_2}, 
$\gamma = (77 \pm 15 \pm 4 \pm 4)^\circ$, 
where the last uncertainty is due to the statistical precision of the CLEO-c
results.

%%%%%%%%%%%%%%%%%%%%%%%%%%%%%%%%%%%%%%%%%%%%%%%%%%%%%%%%%%%%%%%%%%%%%%%%
%%

\section{ADS results}

\begin{figure}[ht]
\begin{center}
\begin{tabular}{ccc}  
\epsfig{file=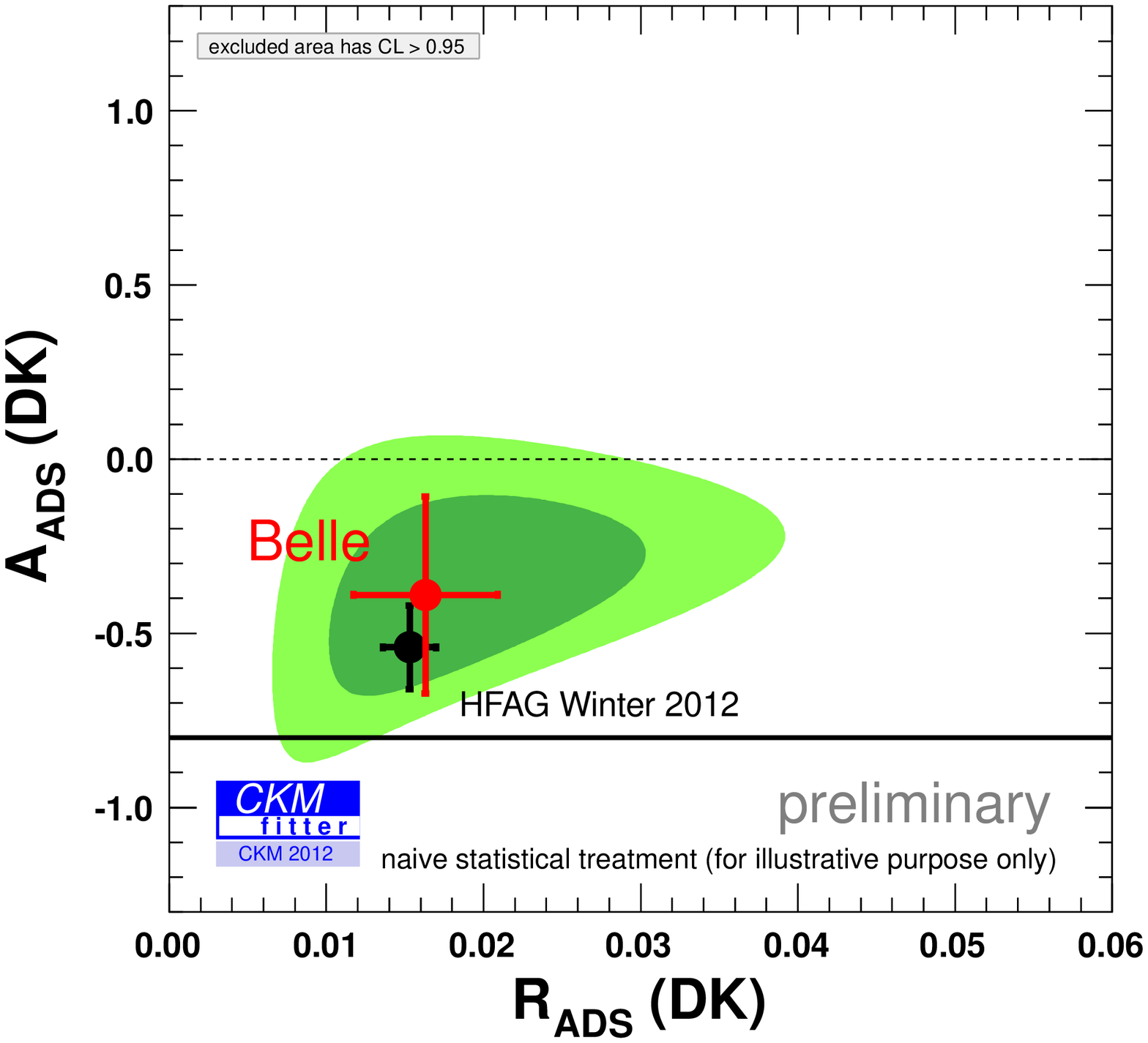,height=1.85in} &
\epsfig{file=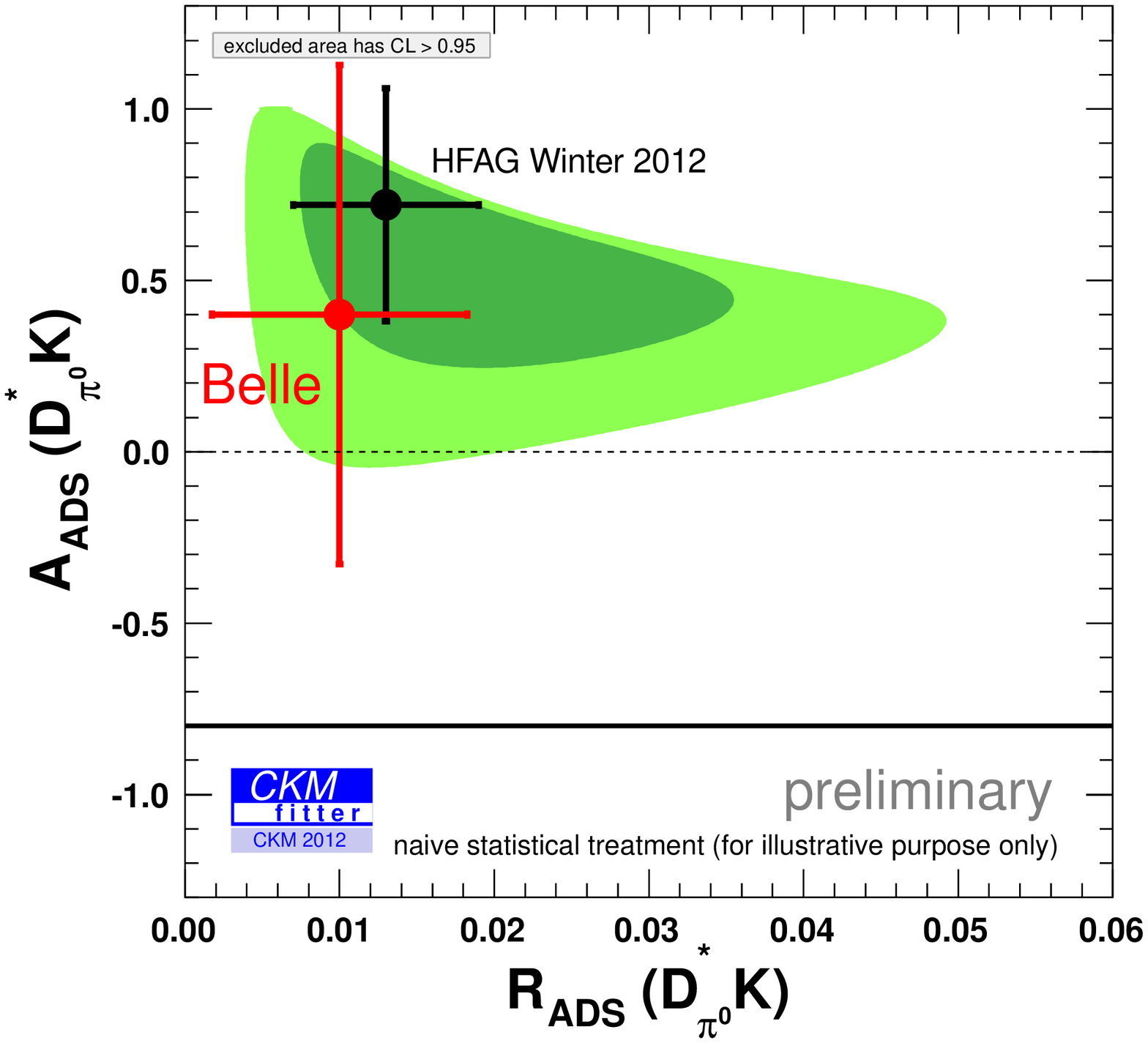,height=1.85in} &
\epsfig{file=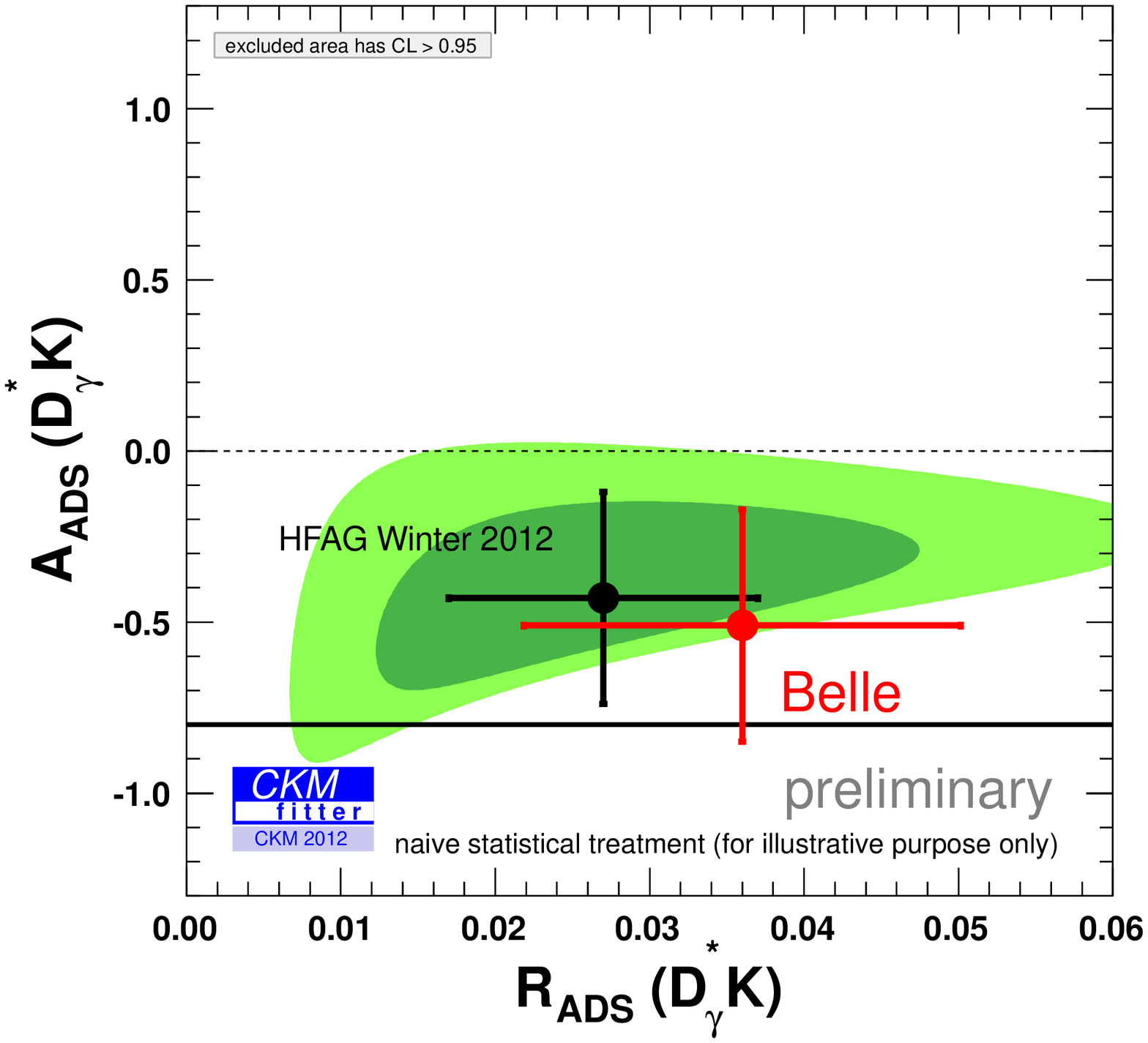,height=1.85in} 
\end{tabular}
\caption{Predictions (from the world averages ($\gamma$, $\delta_B$
and $r_B$) values obtained with the GGSZ method) and 
measurements of the ADS observables for 
$B^\pm \to D K^\pm$ (left), $B^\pm \to D^* K^\pm$ with $D^* \to D \pi^0$ 
(center) and $B^\pm \to D^* K^\pm$ with $D^* \to D \gamma$ (right).}
\label{fig:avsr_ads}
\end{center}
\end{figure}

For the ADS method, Belle has studied the $B \to D^{(*)} K$ decays where
$D \to K^- \pi^+$. The observables measured in the ADS method are the ratio
of the suppressed and allowed branching fractions:
\begin{equation}
{\cal{R}}_{\rm ADS} = 
\frac{\Gamma(B^\pm \to [K^\mp \pi^\pm]_D K^\pm)}{\Gamma(B^\pm \to [K^\pm \pi^\mp]_D K^\pm)}\\
= r_B^2 + r_D^2 + 2 r_B r_D \cos \gamma \cos \delta,
\end{equation}
and
\begin{equation}
{\cal{A}}_{\rm ADS} = \frac{\Gamma(B^- \to [K^+ \pi^-]_D K^-) - \Gamma(B^+ \to [K^- \pi^+]_D K^+)}{\Gamma(B^- \to [K^+ \pi^-]_D K^-) + \Gamma(B^+ \to [K^- \pi^+]_D K^+)}\\
= 2 r_B r_D \sin \gamma \sin \delta/{\cal{R}}_{\rm ADS},
\end{equation}
where $r_D$ is the ratio of the doubly Cabibbo-suppressed and Cabibbo-allowed
$D^0$ decay amplitudes and $\delta$ is the sum of strong phase differences
in $B$ and $D$ decays: $\delta = \delta_B + \delta_D$.
The latest ADS analysis~\cite{Belle_ADS} of $B^\pm \to D K^\pm$ decays with 
$D^0$ decaying to $K^+\pi^-$ and $K^-\pi^+$ (and their charge-conjugated 
partners) uses the full $\Upsilon(4S)$ data sample recorded by the Belle 
experiment. 
The signal yield obtained is $56^{+15}_{-14}$ events, which corresponds 
to the first evidence for an ADS signal (with a significance of 4.1$\sigma$); 
the ratio of the suppressed and allowed modes and asymmetry are summarized 
in Table~\ref{tab_ads}.
\begin{table}[htb]
\caption{Results of the Belle ADS analyses.}
\label{tab_ads}
\begin{center}
\begin{tabular}{lcc}
\hline \hline
Mode & ${\cal{R}}_{\rm ADS}$ & ${\cal{A}}_{\rm ADS}$\\
\hline
$B^\pm \to DK^\pm $ & $0.0163 {}^{+0.0044}_{-0.0041} {}^{+0.0007}_{-0.0013}$ & 
$-0.39 {}^{+0.26}_{-0.28} {}^{+0.04}_{-0.03}$ \\
$B^\pm  \to D^\star K^\pm $, $D^\star \to D \pi^0$ & 
$0.010 {}^{+0.008}_{-0.007} {}^{+0.001}_{-0.002}$ & 
$+0.4 {}^{+1.1}_{-0.7} {}^{+0.2}_{-0.1}$ \\
$B^\pm  \to D^\star K^\pm $, $D^\star \to D \gamma$ & 
$0.036 {}^{+0.014}_{-0.012} \pm 0.002$ & 
$-0.51{}^{+0.33}_{-0.29} \pm 0.08$ \\
\hline \hline
\end{tabular}
\end{center}
\end{table}
The use of two additional decay modes,
$D^* \to D \pi^0$ and $D^* \to D \gamma$, provides an extra handle on the 
extraction of $\gamma$ as explained in Ref.~\cite{Bondar_and_Gershon} 
and illustrated by the predictions of the ADS observables~\cite{CKMfitter} 
from the values of ($\gamma$, $\delta_B$ and $r_B$) obtained with the GGSZ 
method (shown in Fig.~\ref{fig:avsr_ads}). 
This effect (larger ratio for $B^\pm \to D^* K^\pm$ with $D^* \to D \gamma$
and opposite asymmetry between both $B^\pm \to D^* K^\pm$ channels)
is becoming visible in the most recent results from Belle~\cite{Belle_GLW}. 
\begin{figure}[ht]
\begin{center}
\epsfig{file=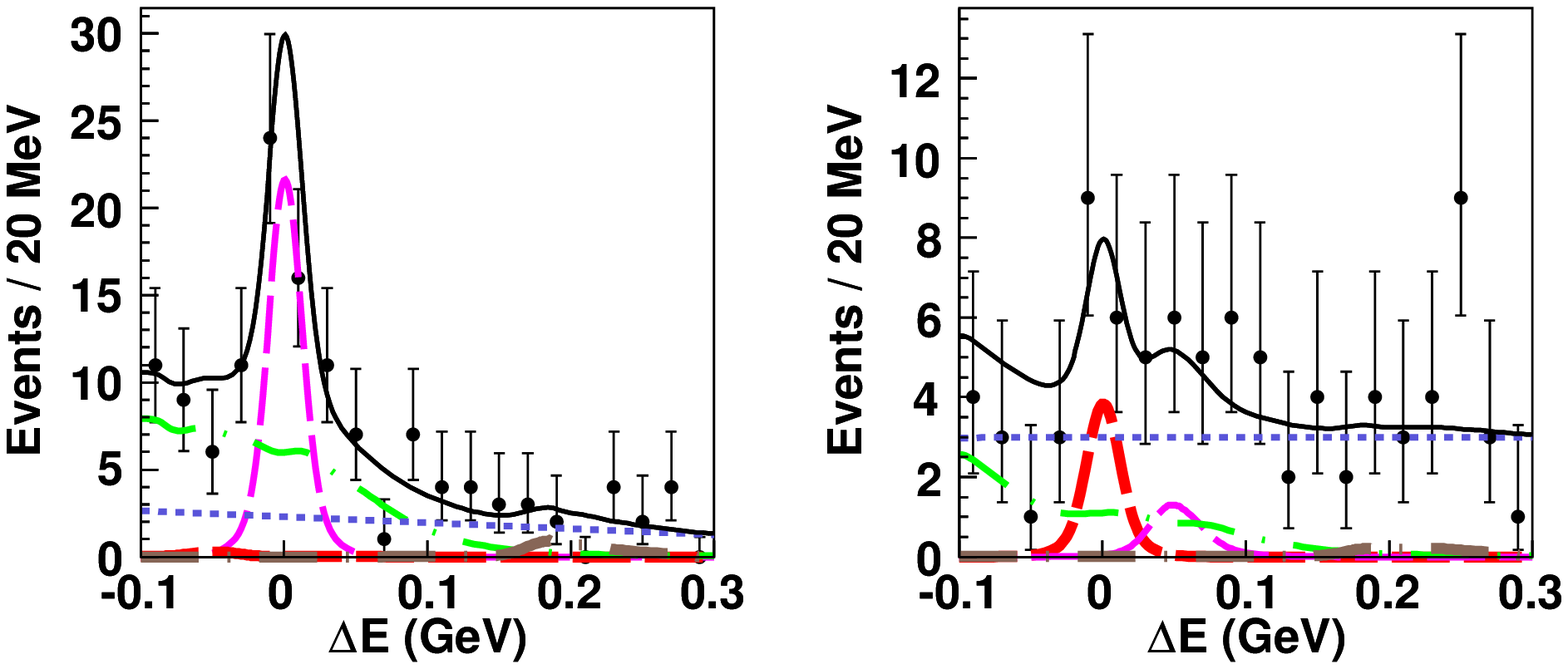,height=1.5in}\\
\epsfig{file=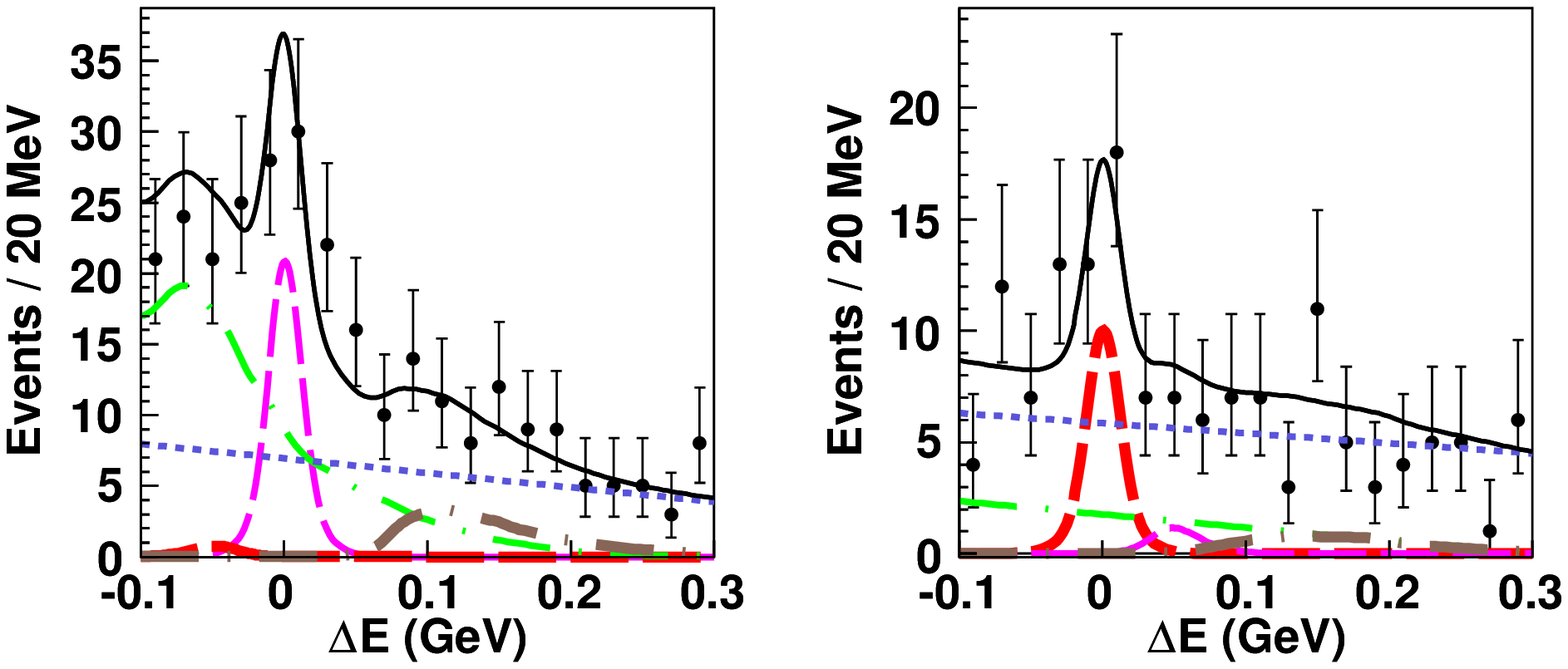,height=1.5in}
\caption{Signal for the $B^\pm \to D^* h^\pm$ decays from Belle ADS analysis.
The plotted variable, $\Delta E$, peaks at zero for signal decays. On the
right plots, $[K^+ \pi^-]_D K^-$ components are shown (by thicker dashed 
curves (red)) for $D^* \to D\pi^0$ (top) and $D^* \to D\gamma$ (bottom).}
\label{fig:dsth_ads}
\end{center}
\end{figure}
%
%%%%%%%%%%%%%%%%%%%%%%%%%%%%%%%%%%%%%%%%%%%%%%%%%%%%%%%%%%%%%%%%%%%%%%%%%
%%
\section{GLW results}

\begin{figure}[ht]
\begin{center}
\begin{tabular}{cccc}  
\epsfig{file=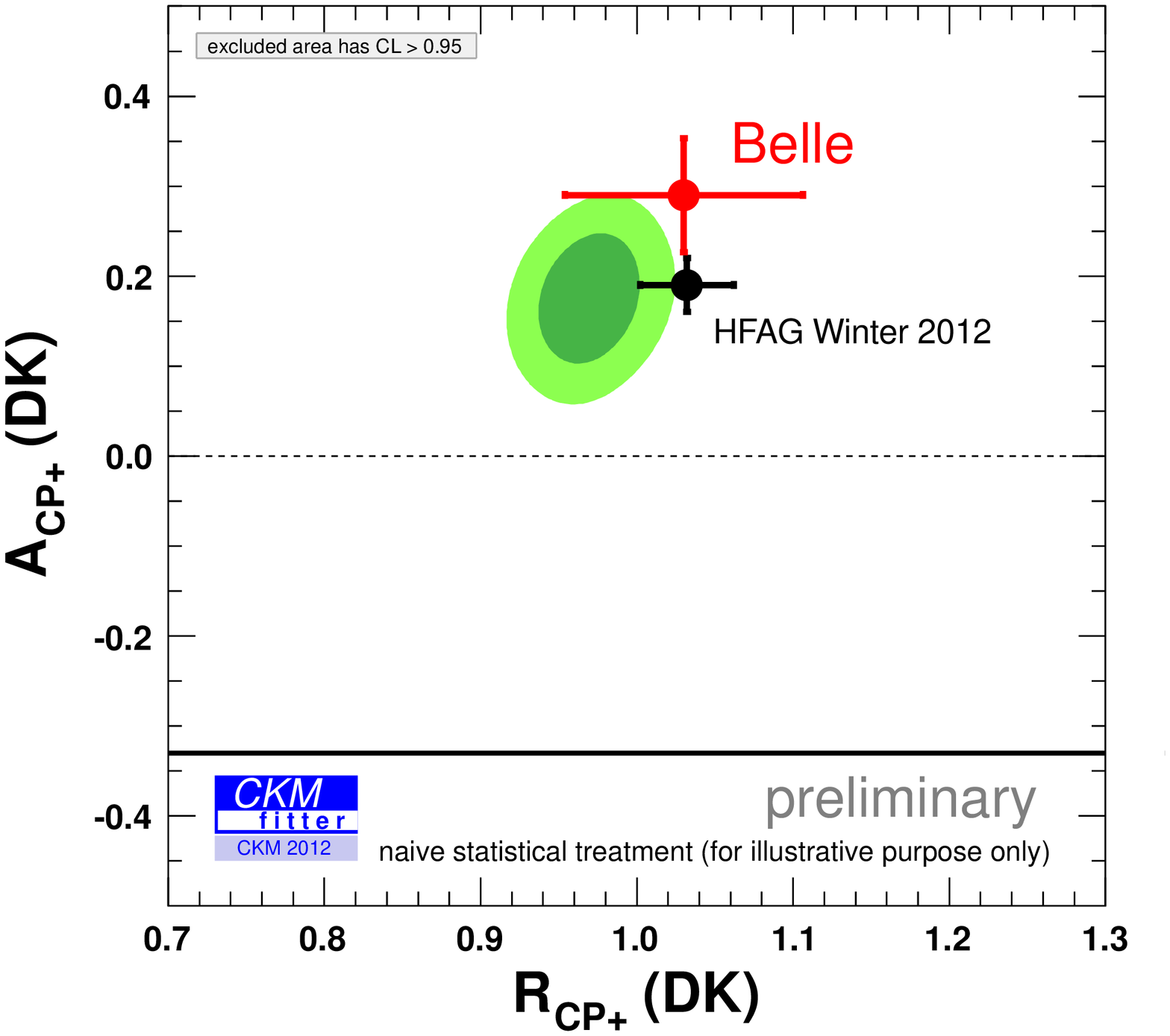,height=1.25in}&
\epsfig{file=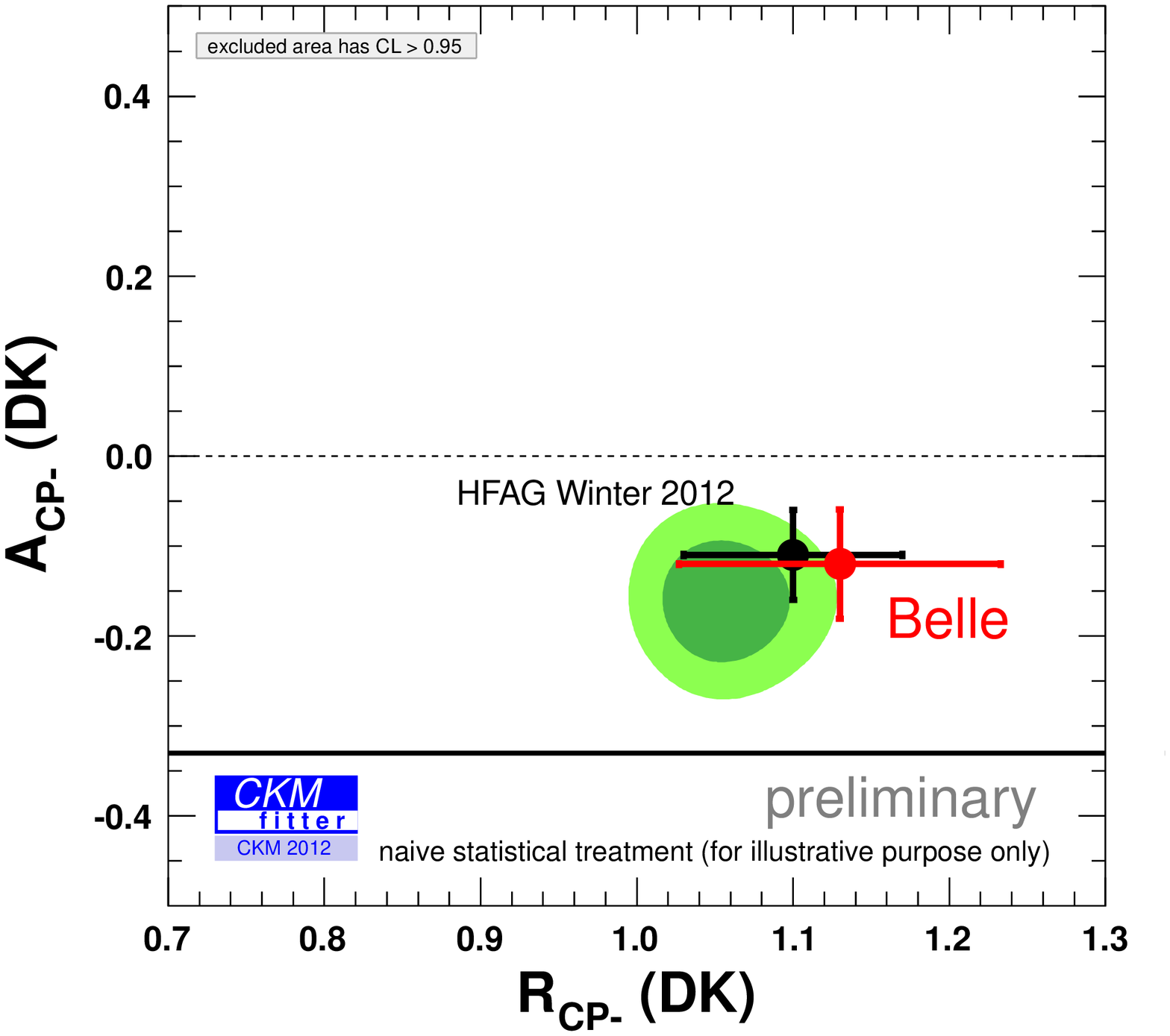,height=1.25in}&
\epsfig{file=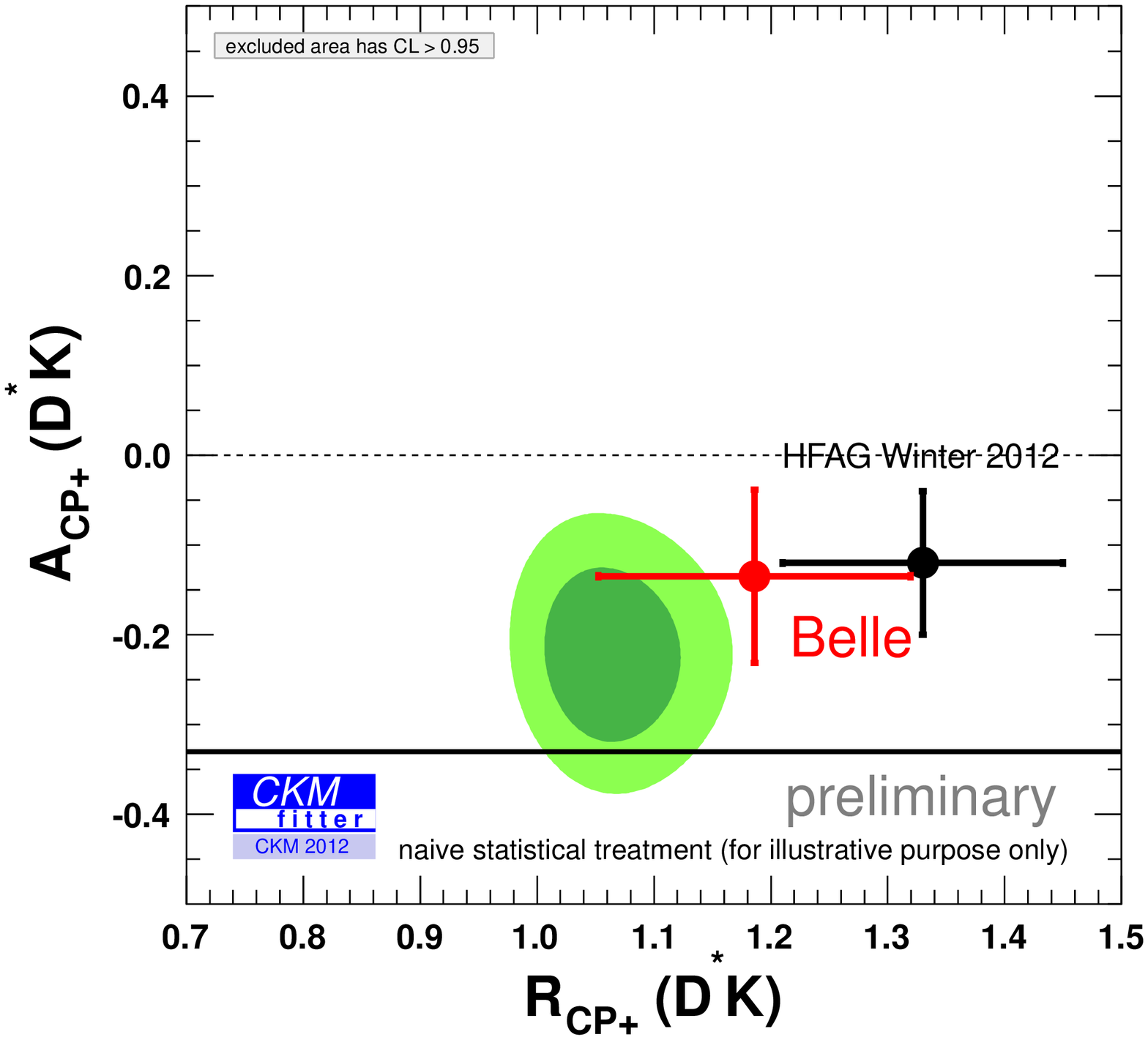,height=1.25in}&
\epsfig{file=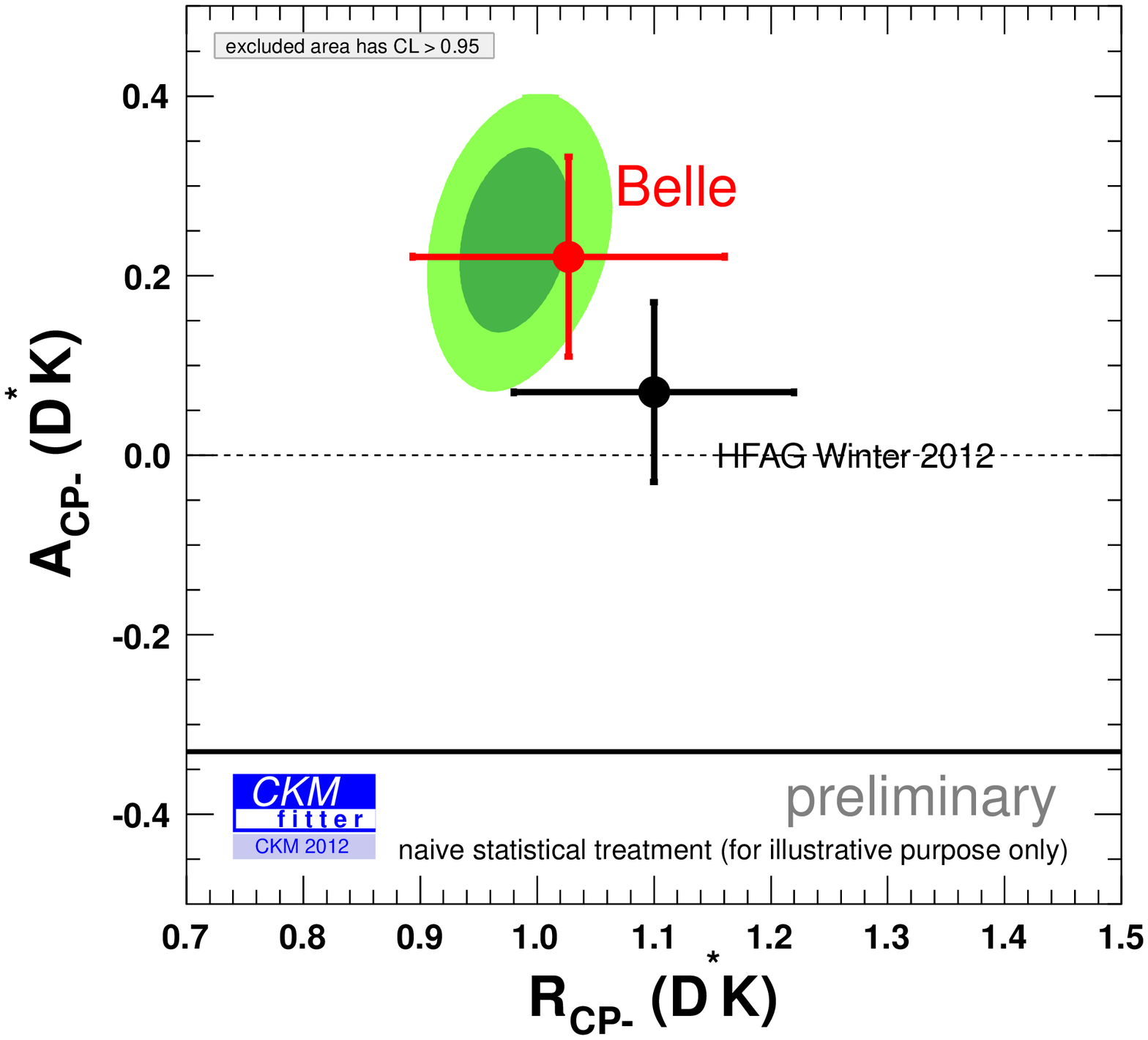,height=1.25in}
\end{tabular}
\caption{Predictions (from the world averages ($\gamma$, $\delta_B$
and $r_B$) values obtained with the GGSZ method) and 
measurements for the GLW observables for 
$B^\pm \to D K^\pm$ (top) and $B^\pm \to D^* K^\pm$ (bottom).}
\label{fig:avsr_glw}
\end{center}
\end{figure}

\begin{figure}[ht]
\begin{center}
\begin{tabular}{c}  
\epsfig{file=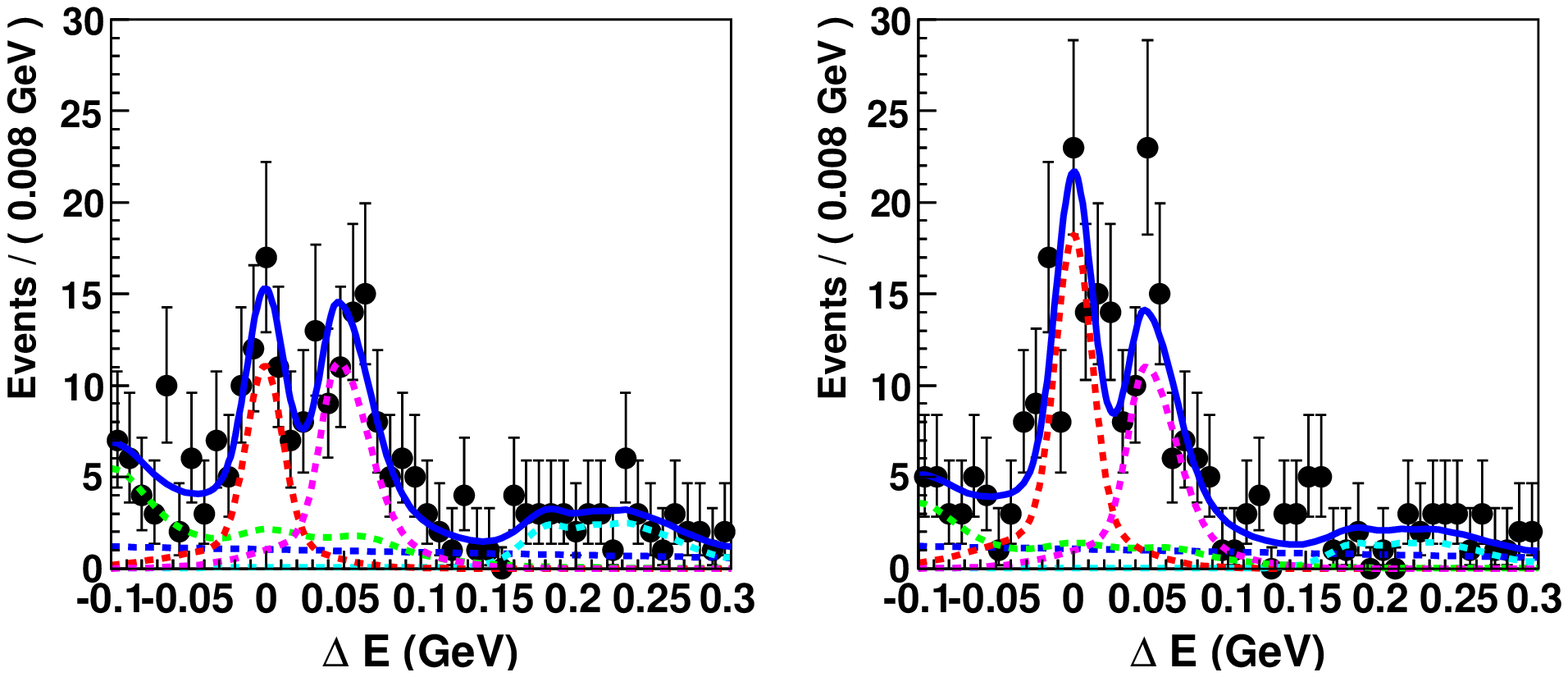,height=1.8in} \\
\epsfig{file=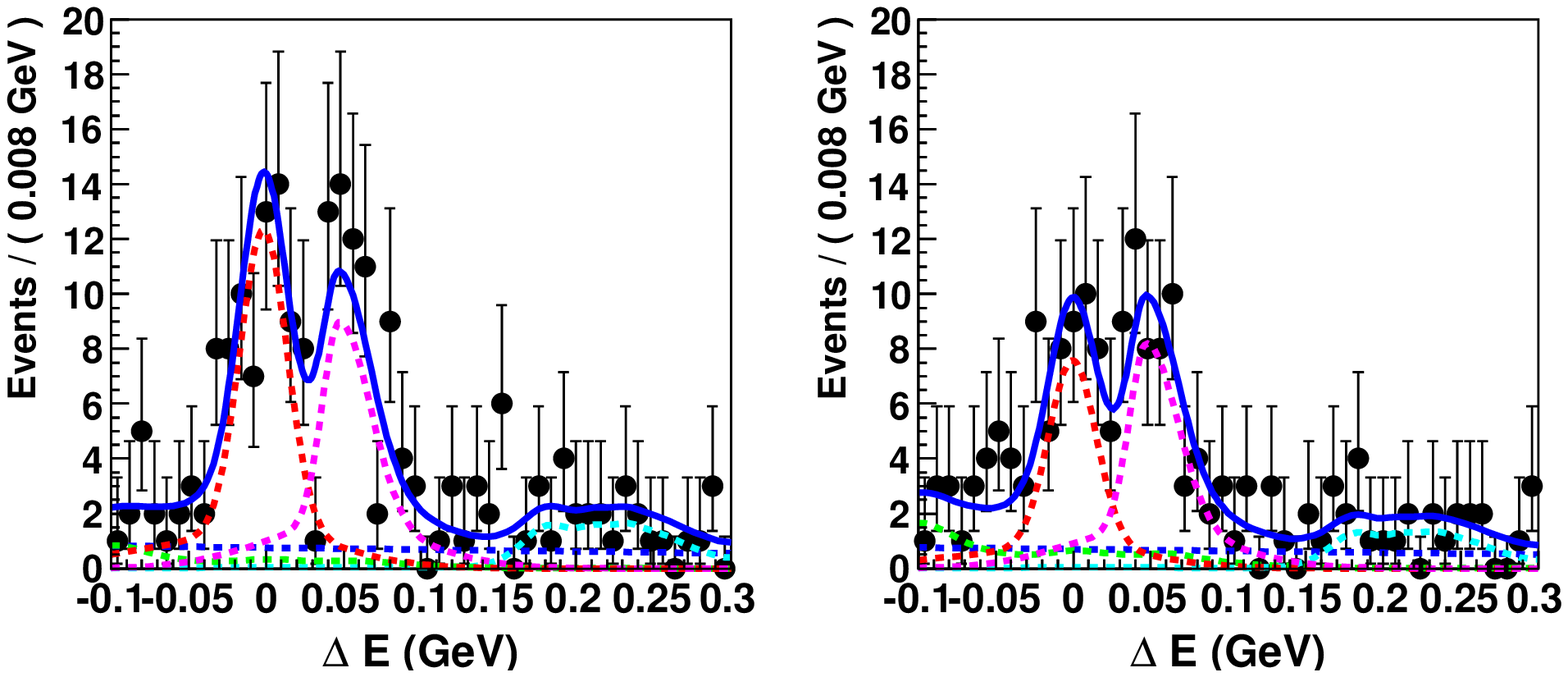,height=1.8in} 
\end{tabular}
\caption{Signals for $B^\pm \to D^* (D\pi^0) K^\pm$ decays: the left 
(right) figures are for $B^-$ ($B^+$) decays and the top (bottom) 
figures are for $\ensuremath{C\!P}$-even ($\ensuremath{C\!P}$-odd) 
eigenstates.}
\label{fig:dsth_glw}
\end{center}
\end{figure}

As alluded earlier, the other interesting class of modes are the ones where
the $D^0$ decays into \CP eigenstates~\cite{GLW} such as $K^+ K^-, \pi^+ \pi^-$ 
($\ensuremath{C\!P}$-even eigenstates) and  $K_S \pi^0$, $K_S \eta$ 
($\ensuremath{C\!P}$-odd eigenstates).
To extract $\gamma$ using the GLW method, the following observables
sensitive to \CP violation are used:
the asymmetries
\begin{equation}
\label{eq:a12}
{\cal{A}}_{CP^\pm} \equiv \frac{\Gamma(B^- \to D_{CP^\pm} K^-) - \Gamma(B^+ \to D_{CP^\pm} K^+)}{\Gamma(B^- \to D_{CP^\pm} K^-) + \Gamma(B^+ \to D_{CP^\pm} K^+)} \\
= \pm\frac{2 r_B \sin \delta_B \sin \gamma }{1 +r_B^2 \pm 2 r_B \cos \delta_B \cos \gamma}
\end{equation}
and the ratios
\begin{equation}
\label{eq:r12}
{\cal{R}}_{CP^\pm} \equiv 2 \frac{\Gamma(B^- \to D_{CP^\pm} K^-) + \Gamma(B^+ \to D_{CP^\pm} K^+)}{\Gamma(B^- \to D^0 K^-) + \Gamma(B^+ \to D^0 K^+)} \\
= 1 +r_B^2 \pm 2 r_B \cos \delta_B \cos \gamma
\end{equation}
Among these four observables, ${\cal{A}}_{CP^\pm}$ and ${\cal{R}}_{CP^\pm}$, 
only three are independent (since ${\cal{A}}_{CP^+} {\cal{R}}_{CP^+} = 
- {\cal{A}}_{CP^-} {\cal{R}}_{CP^-}$).
\begin{table}[ht]
\begin{center}
\caption{Compilation of $R_{\CP}$ and $A_{\CP}$ results for 
$\ensuremath{C\!P}$-even and $\ensuremath{C\!P}$-odd decay modes from Belle. }
\begin{tabular}{llc}  
\hline\hline 
Observables  & $B \to D K$ & $B \to D^* K$ \\[0.3ex] \hline
$R_{\CP^+}$ &  $\phm1.03\pm 0.07\pm 0.03$ & $\phm1.19\pm 0.13\pm 0.03$ \\
$R_{\CP^-}$ &  $\phm1.13\pm 0.09\pm 0.05$ & $\phm1.03\pm 0.13\pm 0.03$\\
$A_{\CP^+}$ &  $+0.29\pm 0.06\pm 0.02$ & $-0.14\pm 0.10\pm 0.01$ \\
$A_{\CP^-}$ &   $-0.12\pm 0.06\pm 0.01$ & $+0.22\pm 0.11\pm 0.01$ \\[0.5ex] 
\hline    \hline   
\end{tabular}
\label{tab_glw}
\end{center}
\end{table}
Recently, Belle updated the GLW analysis using their final data sample of 
$772 \times 10^6$ $B\overline{B}$ pairs~\cite{Belle_GLW}. These results include 
the two $B$ decays: $B^\pm \to D^0 K^\pm$ and $B^\pm \to D^{*0} K^\pm$, where 
$D^{*0} \to D^0 \pi^0$ and $D^0 \gamma$ (the latter modes are shown for the 
first time in this conference, shown Fig.~\ref{fig:dsth_glw}). The signs 
of the ${\cal{A}}_{CP^+}$ 
and ${\cal{A}}_{CP^-}$ asymmetries (Eq.~\ref{eq:a12}) should be opposite (as  
shown by the predictions illustrated in Fig.~\ref{fig:avsr_glw} obtained by 
the CKMfitter group~\cite{CKMfitter}), which is now confirmed by the Belle 
experiment (Table~\ref{tab_glw}).
%
%\clearpage 
%
%%%%%%%%%%%%%%%%%%%%%%%%%%%%%%%%%%%%%%%%%%%%%%%%%%%%%%%%%%%%%%%%%%%%%%%%%%%
%%
\section{$\gamma$ combination from Belle measurements}

%\bigskip

\begin{figure}[ht]
\begin{center}
\begin{tabular}{ccc}  
\epsfig{file=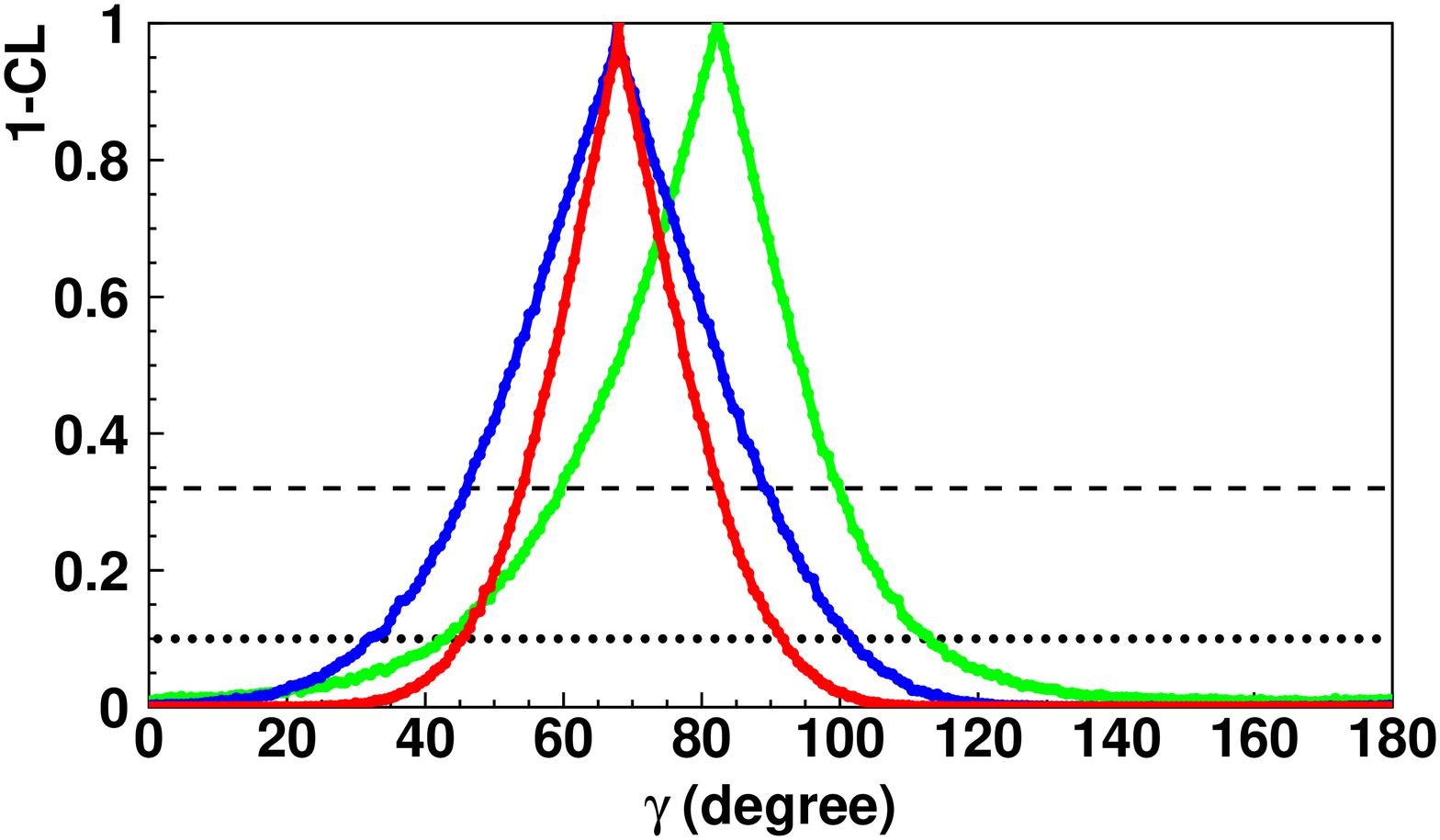,height=1.1in}&
\epsfig{file=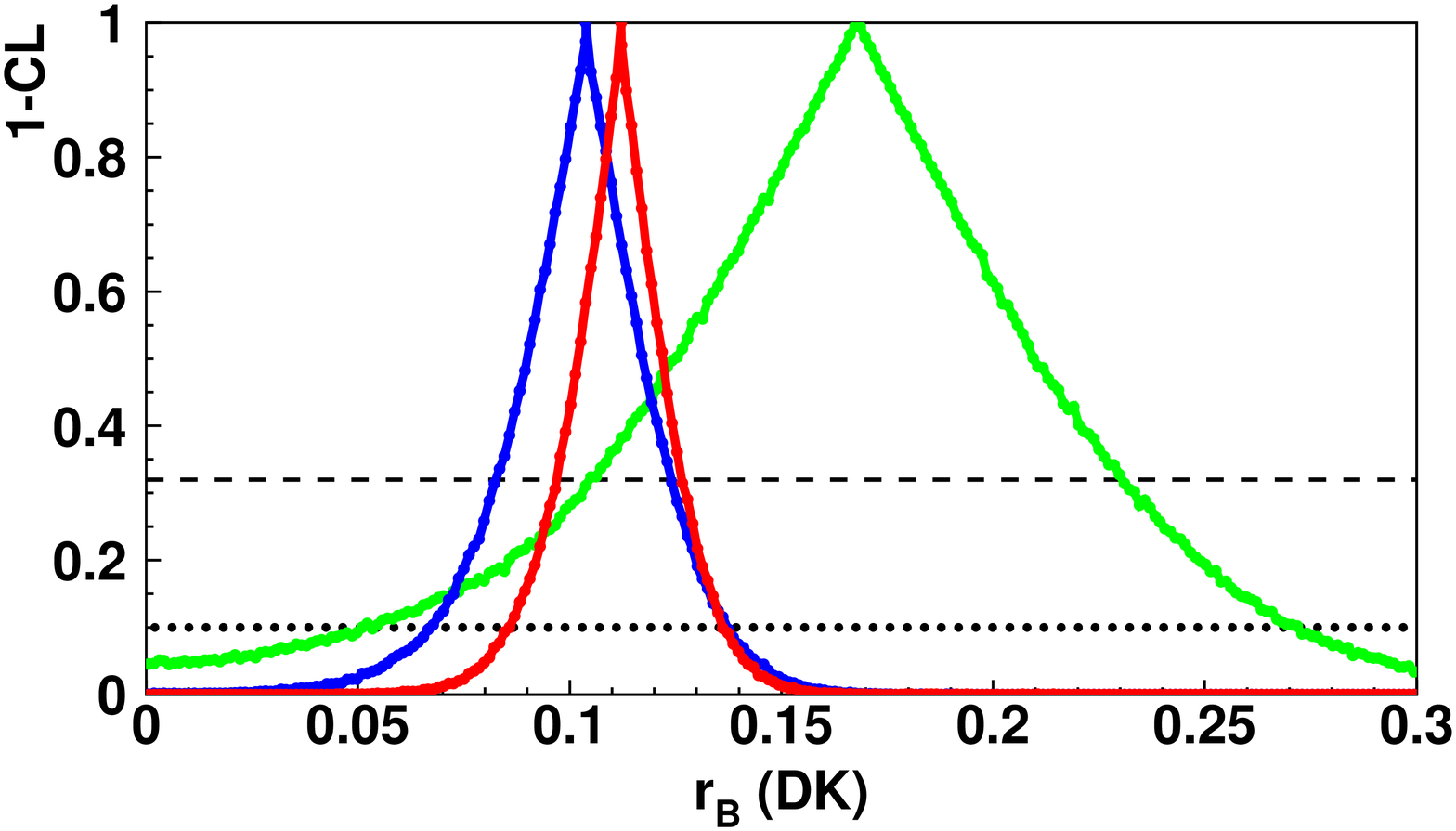,height=1.1in}&
\epsfig{file=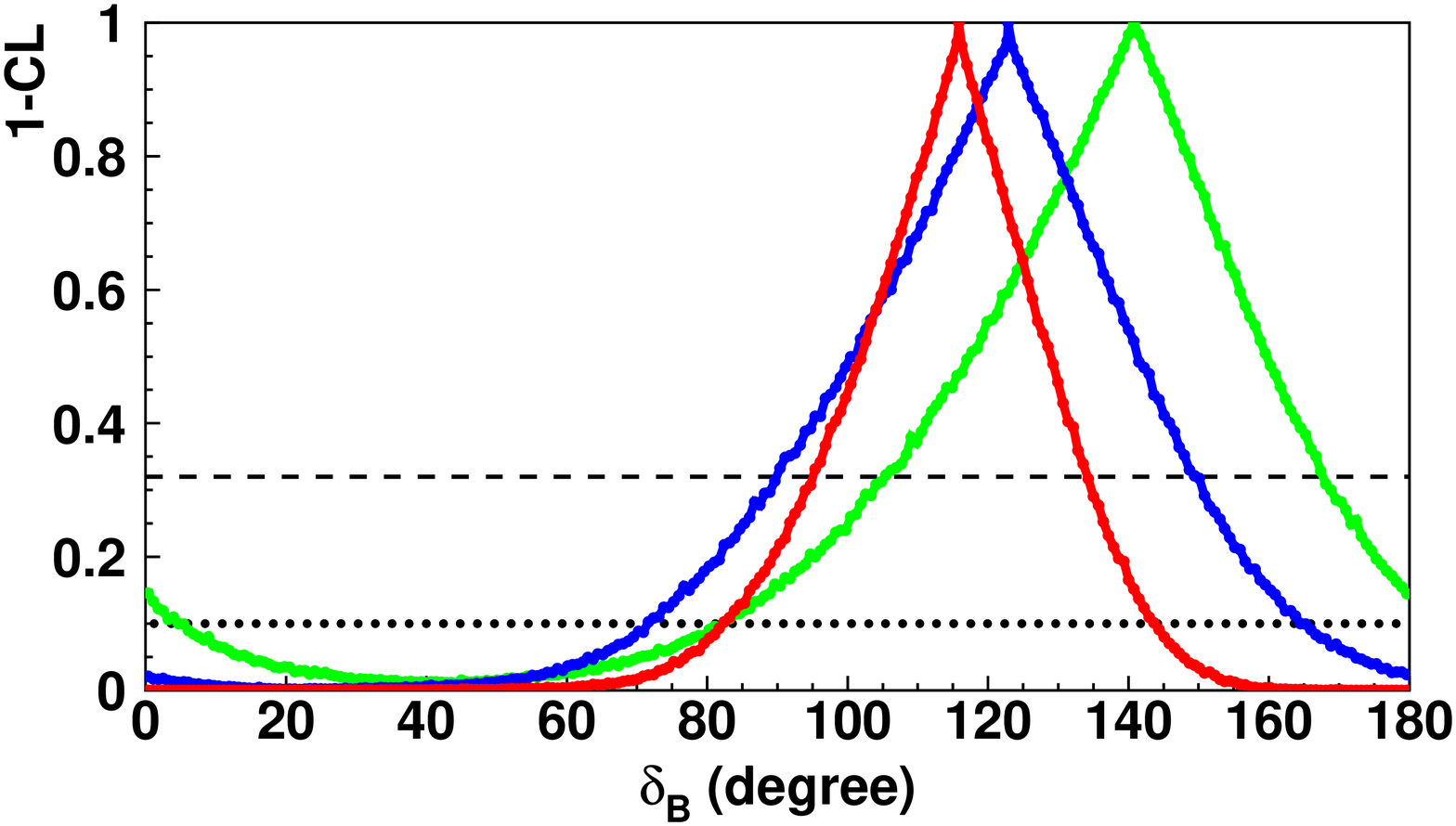,height=1.1in}
\end{tabular}
\caption{1$-$CL curves for $\gamma$ (left), $r_B$ (center) and $\delta_B$ 
(right) from the Belle $D^{(*)}K$ results. The green curve is for the GGSZ 
results, the blue for GGSZ and ADS results using $\delta_D$ from mixing 
and CLEO-c measurements, the red for GGSZ, ADS and GLW results.}
\label{fig:CL}
\end{center}
\end{figure}

We combine the available Belle observables of the $D^{(*)}K$ system obtained
for the GGSZ method (model-dependent results shown in Table~\ref{tab_ggsz}), 
the ADS method (Table~\ref{tab_ads})
and the GLW method (Table~\ref{tab_glw}) using the frequentist procedure also 
exploited in Ref.~\cite{CKMfitter}. The 1$-$CL curves obtained with for 
the angle $\gamma$ as well as for the hadronic parameters ($\delta_B$ and 
$r_B$) of $B \to DK$ mode 
are shown in Fig.~\ref{fig:CL} and the 68\% C.L. intervals are summarized
in Table~\ref{tab_combination}. Belle obtains the most precise $\gamma$ 
measurement to date: $\gamma = (68 {}^{+15}_{-14})^\circ$.
\begin{table}[htb]
\caption{Confidence intervals (68\% C.L.) for the angle $\gamma$ and the 
hadronic parameters of $DK$ ($\delta_B$ and $r_B$) obtained by the combination 
of the $D \to D^{(*)}K$ results of the Belle collaboration.}
\label{tab_combination}
\begin{center}
\begin{tabular}{lccc}
\hline \hline
Method &  $\gamma$ $({}^\circ)$ & $\delta_B$  $({}^\circ)$ &  $r_B$ \\
\hline
GGSZ &         $82 {}^{+18}_{-23}$ & $141 {}^{+27}_{-36}$ & $0.168 {}^{+0.063}_{-0.064}$ \\
GGSZ+ADS &     $68 {}\pm 22$ & $123 {}^{+27}_{-33}$ & $0.104 {}^{+0.020}_{-0.021}$ \\
GGSZ+ADS+GLW & $68 {}^{+15}_{-14}$ & $116 {}^{+18}_{-21}$ & $0.112 {}^{+0.014}_{-0.015}$ \\
\hline \hline
\end{tabular}
\end{center}
\end{table}
%
%\section{Acknowledgments}
%
%\bigskip
%I am grateful to ....

%%%%%%%%%%%%%%%%%%%%%%%%%%%%%%%%%%%%%%%%%%%%%%%%%%%%%%%%%%%%%%%%%%%%%%%%%%%
%\clearpage 

\end{document}

%% file: econfmacros.tex
\def\beq{\begin{equation}}
\def\eeq#1{\label{#1}\end{equation}}
\def\eeqn{\end{equation}}

\def\beqa{\begin{eqnarray}}
\def\eeqa#1{\label{#1}\end{eqnarray}}
\def\eeqan{\end{eqnarray}}

\let\bar=\overbar

\def\Dslash{\not{\hbox{\kern-4pt $D$}}}
\def\dslash{\not{\hbox{\kern-2pt $\del$}}}

\def\msb{{\bar{\ssstyle M \kern -1pt S}}}

%% file: article.bbl
\begin{thebibliography}{99}

%%
%%  bibliographic items can be constructed using the LaTeX format in SPIRES:
%%    see    http://www.slac.stanford.edu/spires/hep/latex.html
%%  SPIRES will also supply the CITATION line information; please include it.
%%

\bibitem{hfag}
Y.~Amhis {\it et al.} (Heavy Flavor Averaging Group),
arXiv:1207.1158 and online update at
http://www.slac.stanford.edu/xorg/hfag.

\bibitem{GLW}
M.~Gronau, D.~London, D.~Wyler, Phys.\ Lett.\ B {\bf 253}, 483 (1991);
M.~Gronau, D.~London, D.~Wyler, Phys.\ Lett.\ B {\bf 265}, 172 (1991).

\bibitem{ADS}
D.~Atwood, I.~Dunietz, A.~Soni, 
Phys.\ Rev.\ Lett. {\bf 78}, 3357 (1997).

\bibitem{GGSZ}
A.~Giri, Yu.~Grossman, A.~Soffer, J.~Zupan, 
Phys.\ Rev.\ D\ {68,2003,054018};
A.~Bondar. Proceedings of BINP Special Analysis
Meeting on Dalitz Analysis, 24-26 Sep. 2002, unpublished.

\bibitem{Belle_GGSZ}
A.~Poluektov {\it et al.} (Belle Collaboration), 
Phys.\ Rev.\ D\ {81,2010,112002}.

\bibitem{GGSZ_2}
A.~Bondar, A.~Poluektov, Eur. Phys. J. C 47 (2006), 347;
A.~Bondar, A.~Poluektov, Eur. Phys. J. C 55 (2008), 51.

\bibitem{CLEOc}
R.~A.~Briere {\it et al.} (CLEO Collaboration), 
Phys.\ Rev.\ D\ {\bf 80}, 032002 (2009).

\bibitem{Belle_GGSZ_2}
H.~Aihara {\it et al.} (Belle Collaboration), 
Phys.\ Rev.\ D\ {\bf 85}, 112014 (2012).

\bibitem{Belle_ADS}
Y.~Horii {\it et al.} (Belle Collaboration), 
Phys.\ Rev.\ Lett. {\bf 106}, 231803 (2011).

\bibitem{Bondar_and_Gershon}
A.~Bondar and T.~Gershon, 
Phys.\ Rev.\ D {\bf 70}, 091503 (2004).

\bibitem{CKMfitter}
J.~Charles {\it et al.} (CKMfitter Group),
Eur.\ Phys.\ J.\ {\bf C41} (2005), 1 and online update at
http://ckmfitter.in2p3.fr/.

\bibitem{Belle_GLW}
Preliminary results presented at Lepton Photon 2011 (BELLE-CONF-1112)
by the Belle collaboration.

\end{thebibliography}
